\documentclass{article}
\usepackage{spconf,amsmath,graphicx}

\usepackage[table]{xcolor}
\usepackage{booktabs}
\usepackage{amssymb,bm}
\usepackage{textcomp}
\usepackage{multirow}

\title{SUT System Description for NIST SRE 2016}
\name{Hossein Zeinali\,$^{1,2}$, Hossein Sameti\,$^1$ and Nooshin Maghsoodi\,$^1$
\thanks{This work was started during Hossein's sabbatical period at Brno University of Technology (BUT) and was continuing at Sharif University of Technology. So, many thanks to BUT for providing setups and systems for running these experiments, especially these guys: Old\v{r}ich Plchot, Pavel Mat\v{e}jka, Luk\'a\v{s} Burget and Jan ``Honza'' \v{C}ernock\'y.}}

\address{
$^1$ Sharif University of Technology, Tehran, Iran \\
$^2$ Brno University of Technology, Speech@FIT and IT4I Center of Excellence, Czech Republic \\
{\small \tt zeinali@ce.sharif.edu, sameti@sharif.edu, nmaghsoodi@ce.sharif.edu}
}

\begin{document}
%
\maketitle
\begin{abstract}
This paper describes the submission to fixed condition of NIST SRE 2016 by Sharif University of Technology (SUT) team. We provide a full description of the systems that were included in our submission. We start with an overview of the datasets that were used for training and development. It is followed by describing front-ends which contain different VAD and feature types. UBM and i-vector extractor training are the next details in this paper. As one of the important steps in system preparation, preconditioning the i-vectors are explained in more details. Then, we describe the classifier and score normalization methods. And finally, some results on SRE16 evaluation dataset are reported and analyzed.
\end{abstract}
\section{Introduction}
\label{sec:intro}

During the past two decades, National Institute of Standards and Technology (NIST) have been organized several speaker recognition evaluations (SRE) and in each of which, they followed one or more challenges. In the current evaluation (i.e. SRE16) the main challenges are:

\vspace{0.5em}\noindent{\bf Mismatch between the training and evaluation datasets:} Because the most of the provided training data is in English, some methods are required for reducing the effects of this mismatch.

\vspace{0.5em}\noindent{\bf Short duration enrollment and test utterances:} This happens more for test utterances where their durations vary from 10 to 60 seconds.

\vspace{0.5em}\noindent{\bf Imbalanced multi-session training:} There are two enrollment conditions for SRE16: three segments available for training some speaker models while only one segment for others.

\section{System description}

In this evaluation, we only used i-vector~\cite{dehak2011front} based systems. Several systems were trained using different features and also different Voice Activity Detections (VADs). All of them used the same Probabilistic Linear Discriminant Analysis (PLDA)~\cite{prince2007probabilistic} back-end.

\subsection{Dataset}

The primary training data is the combination of telephony parts from NIST SRE 2004 - 2008, Fisher English and Switchboard. The unlabeled data from SRE16 development set was used as additional training data. For the final system, we also used labeled data from SRE16 development set. For each subsystem, we used a different subset of these datasets that will be indicated in each section.

\subsection{VAD}

We did experiments with various VAD methods and decided to use two of them. Our main VAD is based on a phoneme recognizer system which trained on Fisher dataset. All frames that recognized as silence or noise were dropped. We will refer to this method by Fisher VAD (FVAD). The secondary VAD is an energy based method that just used in one system. This method is called as Energy VAD (EVAD).

\subsection{Features}

We used four different features. All acoustic features have 19 coefficients along with Energy that makes 20-dimensional feature vectors. Delta and delta-delta coefficients were also used which makes 60-dimensional feature vectors. These features were extracted using a similar configuration: 25\:ms Hamming windowed frames with 15 ms overlap. For each utterance, the features are normalized using short time cepstral mean and variance normalization after dropping the non-speech frames. Three used acoustic features are as follows:
\begin{itemize}
  \item 19 MFCC+Energy
  \item 19 PLP+Energy
  \item Perseus - description of this feature can be found in~\cite{glembek2015migrating}.
\end{itemize}

Beside the acoustic features, an 80-dimensional DNN based Stacked Bottleneck (SBN) feature was used. This feature was trained using Fisher English dataset. The details about SBN can be found in~\cite{matejka2016analysis,zeinali2016deep}.

\subsection{UBM training}

In all systems, a gender-independent diagonal covariance Gaussian Mixture Model (GMM) with 2048 components is used. This model was first trained using about 8000 utterances that were randomly selected from the primary dataset. The relevance MAP with relevance factor 512 was then used for adapting only means of this model by using unlabeled data from SRE16 development set. Doing in this manner was better than adding unlabeled data to UBM training data.

\subsection{i-vector extractor training}

In each system, 600-dimensional i-vectors were extracted from a gender-independent i-vector extractor. This component was trained using about 77000 utterances from the primary dataset and unlabeled data from SRE16 development set. It is worth mentioning again that for UBM and i-vector extractor training only telephony data was used.

\subsection{Preconditioning i-vectors}

\subsubsection{NAP trained on languages}

In order to reduce the effects of mismatch between training and evaluation data languages, we used Nuisance Attribute Projection (NAP) on top of all i-vectors~\cite{dehak2011front,campbell2006svm}. As classes for calculating NAP projection, 20 languages were selected from the primary dataset along with two classes corresponding to major and minor unlabeled data from the development set.

\subsubsection{Regularized LDA}

In addition to NAP projection and prior to training PLDA classifier, i-vectors were first centered by the mean calculated using the primary dataset and then length normalized~\cite{garcia2011analysis}. Based on our previous works on text-dependent speaker verification~\cite{zeinali2016trans,zeinali2016deep}, instead of using conventional Linear Discriminant Analysis (LDA), Regularized LDA (RLDA)~\cite{friedman1989regularized} was used. In this method, the within and between class covariance matrices are calculated using following formulas:
\begin{eqnarray*}
    \mathbf{S}_w & = & \alpha\mathrm{\mathbf{I}} + \frac{1}{S}\sum_{s=1}^{S}\frac{1}{N_s}\sum_{n=1}^{N_s}
    (\mathbf{w_s}^n-\overline{\mathbf{w}_s})(\mathbf{w_s}^n-\overline{\mathbf{w}_s})^t\:, \\
    \mathbf{S}_b & = & \beta\mathrm{\mathbf{I}} + \frac{1}{S}\sum_{s=1}^{S}(\overline{\mathbf{w}_s} - \overline{\mathbf{w}})(\overline{\mathbf{w}_s} - \overline{\mathbf{w}})^t\:,
\end{eqnarray*}
where, $S$ is the total number of classes (i.e. speakers in this paper), $N_s$ is the number of training samples in class $s^{\mathrm{th}}$, $\mathbf{w}_s^n$ is the $n^{\mathrm{th}}$ sample in class $s$, and $\overline{\mathbf{w}_s}=\frac{1}{N_s}\sum_{n=1}^{N_s}\mathbf{w}_s^n$ is the mean of class $s$, $\overline{\mathbf{w}}$ is the mean of total samples, $\mathrm{\mathbf{I}}$ is the identity matrix and $\alpha$ and $\beta$ are two fixed coefficients which must be calculated using the development set.

It is clear that we just add a regularization to each covariance matrix. Alpha and beta parameters are set to 0.001 and 0.01 respectively. Only telephony parts from the primary data were used for RLDA training. The dimension of i-vectors was reduced to 300 by using RLDA.

\subsection{Model enrollment}

We did some experiments on two common schemes of multi-session enrollment: 1) statistics averaging and 2) i-vectors averaging. The second strategy performed slightly better and so we decided to use it for model enrollment with multiple utterances.

\subsection{PLDA}

In all systems, we used PLDA as classifier. The same training data as RLDA is used for PLDA training. The rank of speaker and channel subspaces were set to 200 and 100 respectively.

\subsection{Score normalization}

For score normalization, a specific version of s-norm method was used. In this method, we used trial specific imposter set selection for the t-norm part~\cite{zeinali2017trial} and offline imposter set selection for the z-norm part. For each model during enrollment step, 10000 nearest i-vectors are selected from the primary and unlabeled data. These i-vectors are then scored against the model and the mean and standard deviation of them are used in the z-norm part. For the t-norm part, each test i-vector was first scored against these 10000 i-vectors and then, 5000 largest scores from them were used for calculating mean and standard deviation for the t-norm part. This method is called trial specific because imposter sets are dependent on both model and test i-vectors.

Note that this s-norm method is not symmetric. In the original s-norm method~\cite{kenny2010bayesian}, both imposter sets for the t-norm and z-norm parts are the same and so it is symmetric.

\subsection{Systems}

Our final submission is based on 5 i-vector based systems which are different in input features or VAD as follows:

\begin{itemize}
  \item 60 dimensional MFCC with EVAD
  \item 60 dimensional MFCC with FVAD
  \item 60 dimensional PLP with FVAD
  \item 60 dimensional Perseus with FVAD
  \item 140 dimensional MFCC+SBN with FVAD
\end{itemize}

We did some experiments to find the best strategy for using labeled data from SRE16 development set. When we added this part to RLDA and PLDA training data, we observed a little change in score distributions (i.e. a little shift just on target scores), because the number of speakers in development set (i.e. 20 speakers) compared to training speakers is too few. As a result, we decided to add this data to the training data of these 5 systems and used them as another set for final fusion.

\subsection{Final fusion}

As mentioned in the previous section, we had two sets of 5 systems. In the first one, we didn't add labeled data to training data, but in the second one we did. We trained logistic regression for fusion and calibration of each set of systems using BOSARIS toolkit~\cite{brummer2011bosaris}. SRE16 development trials were used for this fusion training. The final submission is the summation of two fused systems (i.e. with and without labeled data).

\subsection{Execution time and memory consumption}

The reported numbers here were measured using a server with Intel(R) Xeon(R) CPU E5-2640 @ 2.50\:GHz and with 64\:GB memory.

The most consuming steps in our systems are VAD, Extracting features and i-vectors. For acoustic features, the average execution time of these steps using a single thread is about 13 times faster than real time. This number for MFCC+SBN system is about 2.4 times. The memory consumption for these two system types are 3GB and 5GB respectively.

Although the execution time of enrollment and scoring are negligible with respect to the other steps (i.e. it takes about 1.43 second for one model and 1000 test i-vectors), it is worth noting that our score normalization is slower than conventional s-norm. It needs an extra sorting method before selecting scores for calculating mean and standard deviation.

\section{System performance}

We analyze and compare system performance on the SRE16 development data using the equal error rate (EER) and the primary cost. The primary metric in this evaluation is $C_{primary}$, defined as the average cost at two specific points on the DET curve. The detection cost function (DCF) is defined in normalized form as follows:
\begin{equation*}
C_{Norm} = P_{Miss|Tar} + \frac{1-P_{Tar}}{P_{Tar}} \times P_{FalseAlarm|NonTar}\:,
\end{equation*}
where $P_{Target}$ is a priori probability that a trial is a target trial. Actual detection costs will be computed from the trial scores by applying detection thresholds of $log(\beta)$ for the two values of $\beta$, with $\beta_1$ for $P_{Target_1}=0.01$ and $\beta_2$ for $P_{Target_2}=0.005$. And finally the primary cost measure for SRE16 is defined as:
\begin{equation*}
C_{Primary}=\frac{C_{Norm_{\beta_1}}+C_{Norm_{\beta_2}}}{2}\:.
\end{equation*}

Also, a minimum detection cost will be computed by using the detection thresholds that minimize the detection cost.

Table \ref{tbl.results} shows the performance comparison between 5 systems and their fusion for fixed condition as defined in the SRE16 evaluation plan. It is clear that the PLP system works considerably worse than other acoustic features in terms of EER while it performs about the same in terms of minimum $C_{primary}$. This also happens for SBN features concatenated with MFCCs (i.e. MFCC+SBN). This happens because SBN features were trained using Fisher English data and it is proved that the BN features are language dependent and performs the best in trained language. Although this system performs worst, it helps final fusion in terms of both measures.

One interesting observation from this section is the difference between minimum and actual primary cost. It is clear that this difference is not so much and this shows well-calibrated scores without any extra calibration method. This is an important advantage of trial specific imposter set selection for score normalization.

\begin{table*}[th]
  \renewcommand{\arraystretch}{1.3}
  \caption{\label{tbl.results} Performance comparison of different systems and their fusion for SRE16 development set. These results were obtained using NIST scoring script in equalized/unequalized modes. The first section shows results from single systems without any usage of labeled data. The results in the second section were obtained using the same systems from the first section that used labeled data. Contrastive 1 and 2 are the fusion of systems from second and first sections respectively. The last row shows the final submitted system which is the fusion of two Contrastive systems.}
  \vspace{3mm}
  \centerline
  {
  	\setlength\tabcolsep{4pt}
    \begin{tabular}{l l c c c c c c}
      \toprule
      \midrule
      System Name & Features & VAD & Use Labeled Data & Calibration & EER[\%] & $min\:C_{Primary}$ & $act\:C_{Primary}$ \\
      \midrule
      	MFCC\_EVAD		& MFCC		& EVAD 	& No 	& No  & {\bf17.54}\:/\:{\bf17.96} & 0.7355\:/\:0.7295 & 0.7825\:/\:0.7731 \\
        MFCC\_FVAD		& MFCC		& FVAD 	& No 	& No  & 17.75\:/\:18.77 & {\bf0.7192}\:/\:{\bf0.6989} & {\bf0.7612}\:/\:{\bf0.7412} \\
        PLP\_FVAD		& PLP		& FVAD 	& No 	& No  & 19.63\:/\:20.32 & 0.7729\:/\:0.7813 & 0.7978\:/\:0.8063 \\
        Perseus\_FVAD	& Perseus	& FVAD 	& No 	& No  & 17.66\:/\:18.16 & 0.7945\:/\:0.7801 & 0.8109\:/\:0.7986 \\
        MFCC+SBN\_FVAD	& MFCC+SBN	& FVAD 	& No 	& No  & 20.59\:/\:21.93 & 0.7637\:/\:0.7653 & 0.8029\:/\:0.8061 \\
      \midrule
      	MFCC\_EVAD		& MFCC		& EVAD 	& Yes 	& No  & 15.52\:/\:{\bf15.74} & 0.6778\:/\:0.6734 & 0.7329\:/\:0.7207 \\
        MFCC\_FVAD		& MFCC		& FVAD 	& Yes 	& No  & 16.34\:/\:17.34 & {\bf0.6660}\:/\:{\bf0.6539} & {\bf0.7083}\:/\:{\bf0.6925} \\
        PLP\_FVAD		& PLP		& FVAD 	& Yes 	& No  & 18.38\:/\:19.03 & 0.7477\:/\:0.7520 & 0.7557\:/\:0.7634 \\
        Perseus\_FVAD	& Perseus	& FVAD 	& Yes 	& No  & {\bf15.37}\:/\:15.99 & 0.7529\:/\:0.7377 & 0.7834\:/\:0.7623 \\
        MFCC+SBN\_FVAD	& MFCC+SBN	& FVAD 	& Yes 	& No  & 19.43\:/\:20.88 & 0.7408\:/\:0.7413 & 0.7672\:/\:0.7707 \\
      \midrule
        Contrastive 1	& Fusion	& - 	& Yes	& Yes & 13.26\:/\:14.04 & 0.5992\:/\:0.5762 & 0.6170\:/\:0.5889 \\
        Contrastive 2	& Fusion	& - 	& No	& Yes & 15.12\:/\:15.85 & 0.6418\:/\:0.6155 & 0.6648\:/\:0.6334 \\
      \midrule
        Final System	& Fusion	& - 	& Yes	& Yes & 14.14\:/\:14.87 & 0.6196\:/\:0.5978 & 0.6387\:/\:0.6105 \\
      \midrule
      \bottomrule
    \end{tabular}
  }
\end{table*}

The second section of this table reports performance of different systems when labeled data from development set was added to training data. It is obvious that in this case, a large improvement is achieved in terms of EER while the improvement of minimum $C_{primary}$ is not so much. We know that these results are not valid because the systems are overfitted to the development set.

The third section of Table~\ref{tbl.results} shows fusion results on the development set. It is hard to make any conclusion based on these results because we trained logistic regression using this set. Anyway, it is clear that fusion system (i.e. Contrastive 2) performs better in terms of both criteria. Again, the results of Contrastive 1 is not valid  due to overfitting.

The last row of this table shows our final system, which is a simple summation of two systems from the third section. This system is less overfitted to the development set.

\section{Conclusions}

This paper describes SUT system for NIST SRE16. We used different features and VAD in front-end and made the back-end just based on PLDA. Comparison between features showed that acoustic features perform better than bottleneck features in this evaluation due to the language mismatch between training and evaluation datasets. Using NAP was an effective method for reducing the effects of this mismatch. Experimental results proved that using Regularized LDA performs better than conventional LDA for preconditioning i-vectors prior to PLDA training.

For score normalization, we used trial specific imposter set selection method combined with s-norm. This method was the best way for selecting imposter sets. The normalized scores with this method were calibrated well without any additional processing.

Using labeled data from SRE16 development set has a risk of overfitting. So, our final submission system was the fusion of two fused systems with and without using this labeled data to reduce the possibility of overfitting.

\bibliographystyle{IEEEbib}
\bibliography{Speaker,newbib}

\end{document}